\documentclass[runningheads]{llncs}
\usepackage{graphicx}
\usepackage{algorithm}
\usepackage{algpseudocode}
\usepackage{listings}
\begin{document}
\title{Bottom-up Trust Registry in Self Sovereign Identity}
%
%
\author{Kai Jun Eer\inst{1} \and
Jesus Diaz\inst{2}\and
Markulf Kohlweiss\inst{1,2} }
\authorrunning{K. Eer et al.}
%
\institute{University of Edinburgh 
\email{\{keeer,markulf.kohlweiss\}@ed.ac.uk}\and
Input Output Global
\email{jesus.diazvico@iohk.io}\\
}
\maketitle              
\begin{abstract}
Self sovereign identity is a form of decentralised credential management. During credential verification, data exchange only happens between the data owner and the verifier without passing through any third parties. While this approach offers a privacy-centric solution, it poses a challenge. How do verifiers trust that the credential is vouched by a trusted source? More specifically, how do verifiers know that the issuer has the reputation or is authorised to issue the credential? In this paper, we propose a trust registry design that handles the aspect of human trust in self sovereign identity. We also introduce an incentivisation mechanism for the trust registry in order to motivate each stakeholder to participate actively and honestly.

\end{abstract}
\section{Introduction}
In recent years, self-sovereign identity (SSI) as a decentralised identity system has been gaining traction, with the Sovrin blockchain being one of the most notable projects in this area. It aims to return identity data control to the identity owner through cryptographic proofs rather than relying on external identity providers \cite{P5}. However, as there is no central authority regulating the system, SSI faces challenges in key recoverability and trust in identity issuers, which needs to be addressed before SSI can realise its potential. 

The foundation of SSI is verifiable credentials which contain attributes associated with an individual, signed by an issuer. Each credential issuer has an associated decentralised identifier (DID), which is recorded on the public blockchain \cite{P55}. A DID is resolvable into a public key and a service endpoint where interactions with the issuer is communicated. Therefore, while it is cryptographically verifiable that a credential is issued by a specific public-private key pair, there is still a layer of human trust needed. For a verifiable credential to be trusted by a verifier, the verifier needs to trust the issuer of the credential. In an ecosystem of SSI, there could be many valid identity issuers, however the verifier might not recognise all of them. A trust framework that governs the entities which are authorised to issue credentials is critical in managing the layer of human trust. 

\subsection{Contribution} 
The need for a secure and privacy-preserving digital identity system will only increase in the coming years. In this paper, we first analyse current trust frameworks in self sovereign identity, and propose an alternative framework that is more decentralised in nature. 

The main research question of this paper is that in a decentralised setting without a central authority, how to design a trust framework that regulates and incentivises the participation of an ecosystem of identity issuers, which can be trusted by identity verifiers? 

Our main contributions are in two areas of self sovereign identity:
\begin{itemize}
\item
propose a trust registry framework that regulates a web of trust of identity issuers which can be trusted by identity verifiers
\item
propose an incentivisation mechanism in order for the trust registry to be adopted by identity issuers and identity verifiers, while minimising the risk of misbehaviour
\end{itemize}

\section{Related Work} 

During identity verification, the verifier receives a credential presentation proving that the credential is signed by the issuer holding the corresponding DID. However, there might be many identity issuers for the same credential schema, and the identity verifiers should not need to constantly update its local list of all issuers’ DIDs. A verifier might not trust all the issuers too. A trust framework is necessary to govern the human trust aspect in self sovereign identity.  

\subsection{Credential Chaining}

One of the current approaches to SSI trust framework is to use verifiable credentials \cite{P38}. Each valid issuer in the ecosystem holds a verifiable credential that is issued by an issuer one level above it. For example, education institutions hold a credential issued by the government that gives them the rights to issue degrees. During verification, the verifier obtains credential proof presentation by the identity owner. If the verifier does not recognise this level-two credential issuer, it requests another proof presentation from the credential issuer \cite{P13}. The level-two credential issuer then presents its own credential issued by a level-one issuer. This approach is similar to the certificates authority structure \cite{P54} widely adopted in web server authentication. 

The foreseeable practicality challenge of the approach is availability and privacy. To verify credentials, verifiers need to create a communication channel with the issuer(s) in order to climb up the hierarchy and verify credentials. This will only work if the issuers’ agents are always available to prove their own identities. Apart from that, privacy of identity owners decreases as issuers can now correlate the services that the owners access. Issuers gain more control, moving SSI as a user-centric identity model to an issuer-centric model. 

\subsection{Centralised Trust Registry}

A more centralised approach to address the issue of trust in issuers is to have a trust registry that is managed by a trusted authority \cite{P38}. The trusted authority determines which issuers can be listed in the registry. During credential verification, the verifiers query the trust registry, where the trusted authority returns whether the credential issuer is listed in the registry. One drawback of such an approach is that all the verifiers need to trust the single authority to govern the registry, which is often not possible nor desirable in many use cases. If the single authority has biased interest or malicious intent, the trustworthiness of the entire registry will collapse. While there could be multiple trusted authorities that govern the trust registry instead of one, trust is still dependent on a small set of entities.

\section{Bottom-up Trust Framework Design} 

We designed a bottom-up trust framework. There are four direct stakeholders for the trust framework, namely the issuers, verifiers, credential holders and maintainers. The issuers issue credentials to credential holders, who are then able to present these credentials to the verifiers. The maintainers are specific to our trust framework design, where they are tasked to maintain the privacy of the trust registry and facilitate the incentivisation mechanism. 

\begin{figure}[ht]
\vskip 5mm
\includegraphics[width=\columnwidth]{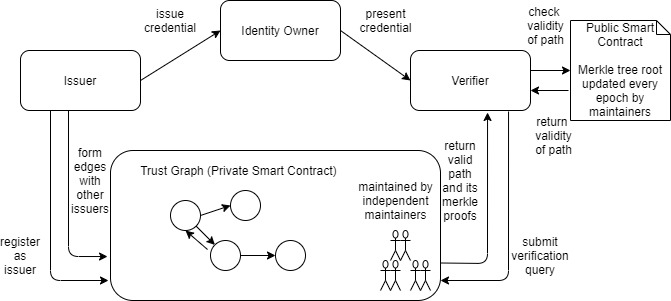}
\caption{System architecture that shows interaction of trust registry with different participants. The trust graph is kept private by utilising a private blockchain maintained by independent maintainers, in order to preserve privacy of issuers and set up an incentivisation mechanism.}
\label{fig:architecture}
\vskip -5mm
\end{figure}

From a high-level view, we use a directed graph of issuers to build a web of trust. Verifiers can compute their level of trust to a particular issuer by traversing the directed graph. To compute the level of trust, verifiers need to pay a fee to a smart contract governing the graph, where the fee will be distributed to issuers. This is to incentivise issuers to issue credentials and form edges with other issuers in the web of trust. An overall protocol architecture is shown in Figure \ref{fig:architecture}.

\subsection{Trust Registry Design}

\begin{figure}[ht]
\vskip 5mm
\centering
\includegraphics[width=50mm]{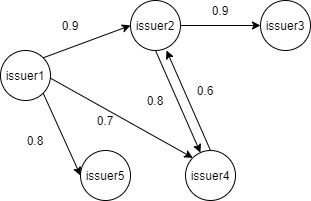}
\caption{An example trust graph that connects the identity issuers. Each edge has a weight which represents the trust score from the source issuer to the destination issuer.}
\label{fig:directed-graph}
\vskip -5mm
\end{figure}

For the bottom-up trust registry, we designed a directed graph that connects identity issuers with other issuers that they trust. An example graph is shown in Figure \ref{fig:directed-graph}. The graph can be represented as G = (V, E) where V is a set of identity issuers represented by their DIDs, while E is a set of directional edges connecting the issuers:
\begin{itemize}
    \item V = \{$v_1$, $v_2$, ……, $v_n$\} where n is the number of issuers listed in the registry.
    \item E = \{$e_{ij}$ ……, \} where $e_{ij}$ is the edge connecting from $v_i$ to $v_j$ and carries a weight of between 0 and 1
\end{itemize}

An edge of $e_{ij}$ means that issuer $v_i$ trusts issuer $v_j$ by a weight, where higher weight indicates higher level of trust. The trust is strictly one-way, therefore $e_{ij}$ is not the same as $e_{ji}$. The value of the weight that is assigned is based on the risk that issuer $v_i$ is willing to take on behalf of issuer $v_j$, which is explained in more details in Section \ref{subsec:economics}.

By implementing the graph as a smart contract, we promise the transparency and immutability of the graph as long as the majority of the blockchain validators are honest. Identity issuers first need to stake an amount of tokens to be listed in the public graph. This is to disincentivise malicious behaviours, where in the event of such behaviour, the involved issuer will lose its stake. Besides that, identity issuers also need to prove in control of the private key corresponding to their DIDs by creating digital signatures. 

\subsection{Privacy Design}

We suggest that the web of trust of identity issuers should be confidential. An identity issuer might want to keep the weight of an edge that it forms with other issuers private, for example, to avoid blackmailing. Therefore, the smart contract which forms the directed graph is deployed on a private and permissioned blockchain. For ease of reference, we will specify the private blockchain validators as \textbf{maintainers} going forward. The maintainers are independent third parties that help to maintain the privacy of the graph and transactions related to it. 

As the graph is computed using a private blockchain run by a small set of maintainers, inevitably it becomes less decentralised. To improve on the trustworthiness, the maintainers have to publish a commitment on the state of the directed graph on a public smart contract at constant timestep. During verification, verifiers are able to compare the partial state returned by the maintainers with the commitment to verify the validity.

The commitment is in the form of a merkle tree root. The commitment is updated at a constant time interval, in which we name the time interval as epoch in this paper. At each epoch, the maintainers collect all the edges E formed in the graph and put them into a merkle tree data structure. Each edge is in the form of ‘source-destination:score’. For example, ‘issuer1-issuer2:0.9’ means that issuer1 trusts issuer2 by a score of 0.9. Once the merkle tree is formed, the maintainers create a multi-signature on the merkle tree root and publish it in a public smart contract.

\subsection{Private Blockchain Relay Server}

The state of a private blockchain is only accessible by its validators, therefore anyone outside the set of validators does not have read and write permission to the blockchain. We introduce the use of relay servers as the bridge between external actors such as issuers and verifiers with the private blockchain. The relay servers are whitelisted as validators in the private blockchain. When external actors call a function in the private smart contract, they create a digital signature using their DID's private key on the function calls and send it to the relay servers. The relay servers then make the function call on their behalf to the private smart contract. This architecture is inspired by the Ethereum gas station network by OpenGSN \cite{P44}.

\section{Credential Verification}

\subsection{Traversing the Graph during Verification}
\label{subsec:traverse}

Each verifier has its own list of trusted issuers. During verification, if the verifier receives a credential proof issued by an issuer it doesn’t recognise, the verifier queries the private blockchain to derive the level of trust. The verifier specifies three arguments, which are its trusted issuers, the credential issuer and a threshold for the trust score. 

Only maintainers have access to the directed graph. When verifiers make a verifcation request (through the relay server), the maintainers return the shortest path in the directed graph that starts from the trusted issuers specified by the verifiers and ends at the credential issuer that is above the threshold score. The path is calculated with a breadth first search algorithm. The trust score is calculated by multiplying the weights of the edges through the path. The reason for using mutiplication is to calculate the transistive trust score, as described in \cite{P50}. If the score is higher than the threshold score specified by the verifier, there is a valid path. If there are two or more paths that have the same path length, the path with the higher trust score is returned. If these paths have the same trust score, a path is randomly picked out of these paths to ensure fairness. 

Recall that at every epoch, maintainers have to publish a merkle tree root commitment of the state of the graph. Once the maintainers have found a valid path, the maintainers generate a proof for each of the edges that forms the path. The maintainers return all the edges that form the path and their corresponding merkle tree proofs. The verifiers are able to check that these edges are accumulated in the merkle tree by verifying the proofs with the merkle tree root stored in the public smart contract. However, there could be privacy concern depending on use cases by simply revealing the edges that forms the path. We propose a alternative described in Section \ref{subsec:zkp}.

\subsection{Economic Incentive Design for Trust Registry Participants}
\label{subsec:economics}

A common challenge for SSI is the lack of incentivisation for issuers. Unlike federated identity providers like Google who collect users behavourial data in exchange for providing free identity service, the privacy by design nature of SSI excludes issuers' involvement during the verification process, removing their source of benefits. 

In our work, we propose that verifiers make a payment \emph{P} to the trust registry when making a verification query. This is logical as verifiers benefit from the SSI ecosystem, where they receive convenience and security when accepting credentials issued by trusted issuers. In traditional identity verification, verifiers usually have to pay the identity issuers directly too. Example of identity issuers are Onfido\footnote{https://onfido.com/} and Jumio\footnote{https://www.jumio.com/}. There are two criterias that we aim to achieve for a trustworthy trust registry:
\begin{itemize}
\item
Issuers are incentivised to issue credentials
\item
Issuers are incentivised to form edges with other issuers in the trust graph, but only with issuers that they truly trust
\end{itemize}

An issuer receives \emph{0.5P} for every verification query where it is the credential issuer, therefore they are incentivized to issue more credentials. When the maintainers return a shortest path in the graph that connects the verifiers' trusted issuers to the credential issuer, all issuers along the path receive an equal share of the remaining \emph{0.5P} paid by the verifier. Thus, issuers are incentivised to form edges with other issuers, as it increases the chances that they are on a valid path during a verification query to the trust registry. Issuers are also incentivised to form an edge with a higher trust score to increase the probability that the path is a valid path with overall trust score higher than the threshold score submitted by the verifier. 

For the trust registry to be trustworthy, we propose a staking mechanism for the issuers as the protocol governance, inspired by \cite{P30}. Issuers have to stake a minimum of \emph{T} tokens in order to be listed in the trust graph. A verifier that makes a verification query to the trust registry is able to challenge the credential issuer if it deems that the credential is issued maliciously. The maintainers then take a vote. If the challenge is successful, the credential issuer loses \emph{T} tokens, while other issuers that are connected directly adjacent to the issuer lose \emph{T * s} tokens based on the trust score \emph{s} that they form with the credential issuer. The stake tokens lost by the issuers are distributed to the verifier as compensation. With this mechanism, issuers are able to quantify the amount of risk they are willing to take on other issuers based on the trust score \emph{s} they assign \cite{P36}. 

When the issuers lose some stake tokens, its staking token balance in the trust registry might go below the minimum \emph{T} tokens required. These issuers need to top up the number of staking tokens to meet the minimum requirement, otherwise the maintainers remove the issuers from being a node in the trust graph. In such a scenario, the issuers will not earn any more rewards from future verification queries. 

\section{Implementation}

We use a proof-of-authority (PoA) blockchain for the private blockchain, where each blockchain validator needs to have its identity verified before being allowed to participate. We set up an EVM private blockchain using the Hyperledger Besu framework to prototype our proposed design. The reason for using Hyperledger Besu is its capabilities to form a privacy group, where smart contracts in the privacy group and their corresponding transactions are protected from the public \cite{P43}. This private blockchain is used to maintain the state of the directed graph. In our work, we use four independent maintainers as the validators.

We use the Kovan blockchain (an Ethereum testnet) as the public blockchain. The smart contract Registry.sol which stores the directed graph is deployed on the private blockchain. The smart contract MerkleProof.sol which records the merkle tree root is deployed on Kovan and can be viewed on a block explorer such as Kovan Etherscan\footnote{https://kovan.etherscan.io/address/0xc4516FD2317210a278da78F06f7F1CeBC495A4a8}. 

\section{Proposed Improvements} 
\label{subsec:zkp}

We explore the use of zero knowledge proofs \cite{P49} to reduce the trust needed on maintainers and limit the revealed information to the verifiers. Our current protocol relies on the maintainers to reveal a valid path and its corresponding merkle proofs when a verifier makes a verification query. The protocol also relies on the maintainers to distribute the verification fees to relevant issuers. 

With the use of non-interactive zero knowledge (NIZK) proofs, maintainers are able to prove that they distribute the verification fee honestly. The proposed protocol is as follows, although we have not analysed this in details and is left as future work:
\begin{enumerate}
\item
Maintainers keep a record of the balances of each issuer. The balances are stored in a merkle tree data structure, and the merkle tree root is recorded on a public smart contract.
\item 
The verifier pays the verification fee to a public smart contract, and subsequently makes a verification query to the maintainers, stating the trusted issuer, the credential issuer and the threshold trust score in order for the verification to be accepted.
\item
The maintainers first check that the verifier has made the payment to the public smart contract. Then, the maintainers generate a NIZK proof to prove the existence of a valid path between the trusted issuer and credential issuer. The maintainers update the balances of issuers that are part of the valid path, giving rise to a new merkle tree. The NIZK proof also encapsulates that the balances of issuers are updated correctly. The maintainers update the new merkle tree root on the public smart contract and also publish the zero knowledge proof. 
\item
Time is divided into epochs, for example an epoch could be one month. During that epoch, steps 2 and 3 are repeated whenever a verifier makes a verification query. 
\item
At the end of the epoch, the maintainers reveal the balances of all the issuers. The correctness of these balances can be verified through the merkle tree root and the NIZK proofs generated previously. 
\item
Once the balances are revealed, issuers are able to claim their corresponding payments from the public smart contract. 
\end{enumerate}

The program used to generate the zero knowledge proof is described in Algorithm \ref{alg:zkp}, where it first verifies that these edges are accumulated in the graph merkle tree, and lastly verifies that the balances of the issuers are updated correctly based on the edges. 

\begin{algorithm}
\caption{Generate zero knowledge proof on existence of valid path and correct update of issuers' balance}
\begin{algorithmic}
\Require $edges$, $balances$, $proof_{edges}$ \Comment{private inputs}
\Require $source$, $destination$, $root_{graph}$, $root_{balance(previous)}$, $root_{balance(updated)}$, $threshold$ \Comment{public inputs} 
\State $P_{cred} \gets 0.5P$ \Comment{credential issuer gets half of the verification fee}
\State $P_{others} \gets 0.5P / edges.length$ \Comment{other issuers along the path gets the remaining half}
\State verify that $edges$ is a valid path from $source$ to $destination$ and above $threshold$
\State verify $balances$ corresponds to $root_{balance(previous)}$
\For{$edge$ in $edges$}
\State verify $edge$ accumulated in $root_{graph}$ by comparing against $proof_{edges}$
\State $issuer \gets edge.issuer$ \Comment{get the destination issuer in the edge}
\If{$issuer$ is credential issuer}
  \State add $P_{cred}$ to its balance 
\Else
  \State add $P_{others}$ to its balance
\EndIf
\EndFor
\State verify that updated $balances$ corresponds to $root_{balance(updated)}$
\end{algorithmic}
\label{alg:zkp}
\end{algorithm}

With this proposed use of zero knowledge, we are able to reduce the trust on maintainers. If the maintainers coerce and update the issuers’ balance maliciously, it is detectable as the correctness of the zero knowledge proof will be false. We maintain the privacy of the verification queries as the balances of the issuers are aggregated and only revealed at the end of an epoch, therefore it is difficult to correlate verifications to issuers in the directed graph, provided there is a large enough number of issuers in the graph \cite{P57}. The tradeoff to this approach is the reduction in scalability, as maintainers need to update the merkle tree root of the balances in the public smart contract for each verification query. Generating non-interactive zero knowledge proofs are also costly. 




\section{Conclusion}

Self sovereign identity (SSI) is an emerging field in this time where the world is undergoing rapid digital transformation. It offers a privacy-centric solution where users are always in control of their data, in terms of what data they consent to sharing and with whom. However, it is important to understand that human trust is still the underlying factor under the technology. Anyone is able to vouch for the correctness of a particular piece of data, however the trustworthiness of the vouched data depends on the reputation of the issuer. In a multi-party system where multiple issuers and verifiers exist, there is a need for a trust registry to manage the human trust layer of these different entities. 

Our work introduces a bottom-up approach for a self sovereign identity trust registry using the web of trust model. For the trust registry to be adopted, we also introduce an incentivisation mechanism such that all direct stakeholders of the trust registry are incentivised to perform proactively and honestly. In comparison to the current common design and protocols for SSI trust registries, we are able to offer an alternative method, which can be useful in certain use cases where there is no single source of truth to verify the validity of the credential issuers. 

\subsection{Future Work}

As mentioned in Section \ref{subsec:zkp}, the zero knowledge extension to our protocol is not fully experimented and evaluated yet. Besides that, our protocol currently relies on the maintainers to preserve the privacy of the directed graph. It is possible to use the technique of homomorphic encryption \cite{P56} in cryptography such that the directed graph is stored in an encrypted decentralised storage, and changes to the state of the graph can be made without revealing the graph itself. This makes it harder for a malicious maintainer to reveal the directed graph outside of the protocol, however the tradeoff is scalability. Other possible approaches worth exploring further include using secure multi party computation \cite{P40} to compute the trust score in the directed graph, removing the need for maintainers to answer verification queries. Related work is \cite{P34}.

Our incentivisation mechanism relies on the verifiers paying a small fee to make a verification query. It is worth exploring different economic interactions such as the credential holder being the fee-paying entity. A game theory approach to validate the incentivisation mechanism will define clearer parameters too. 

%
%

%
%
%
\bibliographystyle{splncs04}
\bibliography{mybibfile}
%

\end{document}